\documentclass[10pt,letterpaper,twocolumn]{article} 
\usepackage{ol2,multicol}
\usepackage[draft]{hyperref}
\usepackage{amsmath}

\newcommand{\ket}[1]{|#1\rangle}

\newcommand{\mrm}{\mathrm}

\newcommand{\ii}{\mrm{i}}

\usepackage{color}

\begin{document}
\twocolumn[
\title{Three-photon electromagnetically induced transparency \\ using Rydberg states}
\author{Christopher Carr$^1$, Monsit Tanasittikosol$^1$, Armen Sargsyan$^2$, David Sarkisyan$^2$, \\Charles S. Adams$^1$ and Kevin J. Weatherill$^1$}
\address{$^1$Department of Physics, Durham University, South Road, Durham, DH1 3LE, UK\\
$^2$Institute for Physical Research, Armenia National Academy of Science, 0203 Ashtarak-2, Armenia}
\email{k.j.weatherill@durham.ac.uk} 


\begin{abstract} We demonstrate electromagnetically induced transparency (EIT) in a four-level cascade system where the upper level is a Rydberg state. The observed spectral features are sub-Doppler and can be enhanced due to the compensation of Doppler shifts with AC Stark shifts. A theoretical description of the system is developed which agrees well with the experimental results and an expression for the optimum parameters is derived. \end{abstract}

\ocis{020.1480, 020.1670, 020.5780, 300.6210}]


Non-linear optical effects in ensembles of Rydberg atoms is a topic of burgeoning interest \cite{prit12}. The large polarizability of the Rydberg states leads to very large DC electro-optic effects \cite{moha08} and the strong, long-range interactions between neighboring atoms have been exploited to demonstrate cooperative enhancement of non-linearities \cite{prit10,petr11} and single photon generation \cite{dudi12}. Rydberg non-linear optics offers the prospect of non-local effects \cite{sevi11}, photon-photon interactions \cite{gors11} and single photon subtraction \cite{hone11}.

Multi-photon excitation schemes are of particular interest because they can eliminate motional dephasing of dark state polaritons \cite{bari12} and allow Rydberg states to be accessed using inexpensive and convenient diode laser systems. Recent experiments have demonstrated Rydberg spectroscopy based on population transfer \cite{thou09,john12}, and similar coherent processes have been studied theoretically \cite{ryab11}.

\begin{figure}[b]
\centerline{{\includegraphics[width=3.38in]{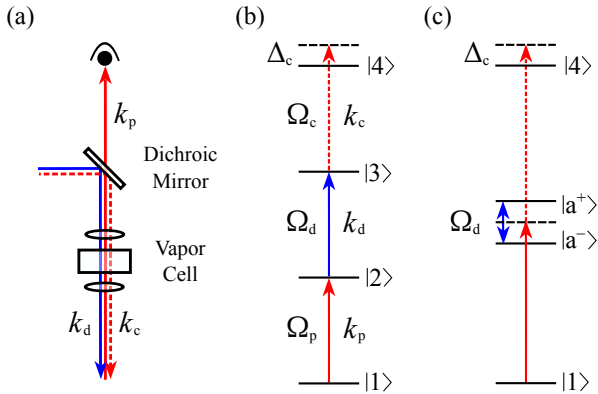}}}
\caption{a) Experimental setup. b) Bare state atomic energy-levels.  c) Dressed state energy levels; the dressing laser acts as a perturbation to the system.}
\end{figure}

Here we study, experimentally and theoretically, coherent three-photon electromagnetically induced transparency (EIT) \cite{flei05} involving Rydberg states in Cesium vapor. A schematic of the experimental setup and the atomic energy level scheme are shown in Fig. 1 (a) and (b) respectively. States $|1\rangle$, $|2\rangle$, $|3\rangle$ and $|4\rangle$ represent the 6S$_{1/2}$, 6P$_{3/2}$, 7S$_{1/2}$ and 26P$_{3/2}$ states respectively. A weak 852~nm probe beam coupling states $|1\rangle$ and $|2\rangle$, is focused through the center of a 2.4~mm Cs vapor cell and then detected at a photodiode. The counterpropagating 1470~nm dressing beam is stabilized to the $|2\rangle \rightarrow |3\rangle$ transition using excited-state polarization spectroscopy \cite{carr11}. The 790~nm coupling beam also counterpropagates with the probe and excites the $|3\rangle \rightarrow |4\rangle$ transition. We scan over either the $|1\rangle \rightarrow |2\rangle$ or $|3\rangle \rightarrow |4\rangle$ transition whilst the other two lasers are on resonance. The 1/e$^2$ radii of the probe, dressing and coupling beams are 14~$\mu$m, 30~$\mu$m and 20~$\mu$m, the wavevectors are $k_{\rm p}$ , $k_{\rm d}$ and $k_{\rm c}$, the beam powers are $P_{\rm p}$, $P_{\rm d}$ and $P_{\rm c}$ and the Rabi frequencies are $\Omega_{\rm p}$, $\Omega_{\rm d}$ and $\Omega_{\rm c}$ respectively. The temperature of the cell is maintained at 50$^\circ$C giving a number density, $N_0$, of 2.0~x~10$^{11}$~cm$^{-3}$.

\begin{figure}[t]
\centerline{{\includegraphics[width=3.0in]{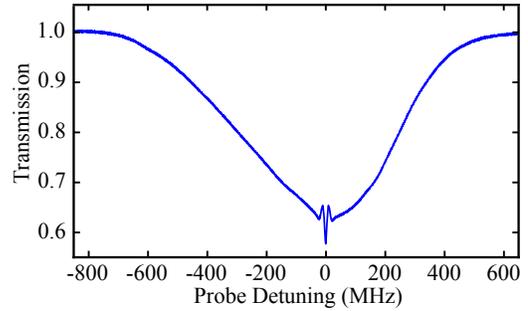}}}
\caption{Experimental weak probe transmission spectrum for $P_{\rm p}=20$~nW, $P_{\rm d}=500$~$\mu$W and $P_{\rm c}=200$~mW.}
\end{figure}

\begin{figure*}[t]
\centerline{{\includegraphics[width=6in]{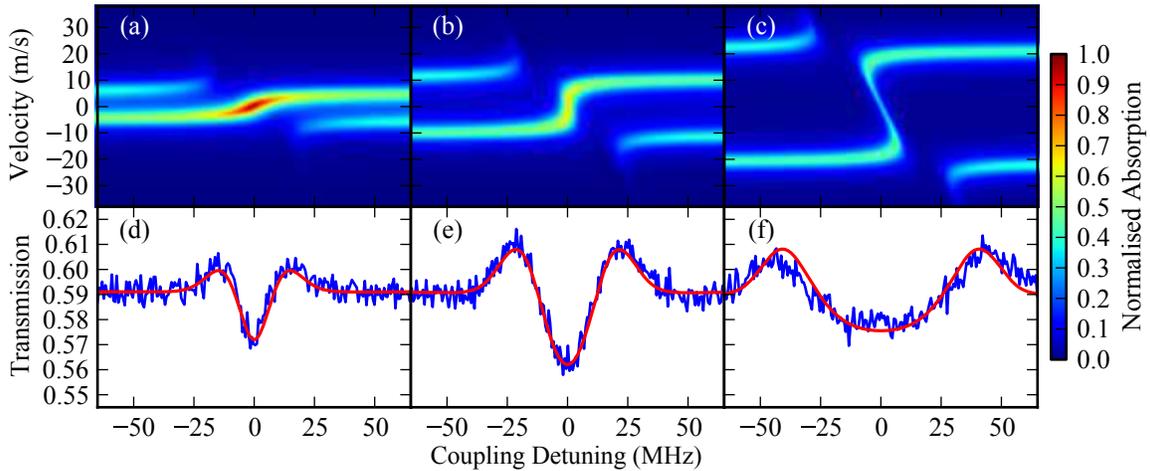}}}
\caption{(color online) Top: Absorption coefficient per velocity classes from the four-level model for different values of $\Omega_{\rm d}/\Omega_{\rm c}$, (a) 0.6 (b) 1.2 and (c) 2.5. At the optimum ratio of 1.2 (calculated from Eq. (\ref{eq:eq7})), the bright state is insensitive to the atomic velocity. Bottom: Typical experimental spectra with fits from the full theoretical model for different values of dressing laser power P$_{\rm d}$, (d) 17.4~$\mu$W, (e) 79.7~$\mu$W and (f) 345.2~$\mu$W. Experimental parameters fixed at P$_{\rm p}$ = 20~nW, P$_{\rm c}$ = 41.4~mW and $\Delta_{\rm p}=\Delta_{\rm d}=0$~MHz.}
\end{figure*}

A typical spectrum obtained when the probe frequency is scanned is shown in Fig. 2. A narrow spectral feature is observed within the D$_2$ Doppler absorption lineshape with enhanced absorption on resonance and enhanced transmission either side of resonance. This lineshape can be understood by considering the dressed state picture shown in Fig. 1 (c) where the system now contains two cascade EIT systems \cite{armen10}. The result is two dark states which overlap in frequency space and interfere constructively to give enhanced or electromagnetically induced absorption (EIA)\cite{McGlo01}. When the coupling laser is scanned instead, a similar lineshape is observed but on a flat transmission background as shown in Fig. 3 (d-f).

To model the system we consider the four-level system shown in Fig. 1 (b). Using semi-classical theory and including the Doppler effect \cite{tana12}, the complex susceptibility of the system per velocity class in the weak probe approximation is given by
\begin{equation}
\chi(v){\rm d}v=N(v){\rm d}v\frac{{\rm i}d_{21}^2}{\epsilon_0\hbar }\left[\frac{\Gamma_2}{2}+\ii k_{\rm p}v+\frac{\ii\Omega_{\rm c}}{2}\frac{\rho_{31}}{\rho_{21}}\right]^{-1},
\end{equation}
where $d_{21}$ is the probe dipole matrix element and $N(v){\rm d}v$ is the atomic density per velocity class $v$. This Gaussian distribution is written in terms of the most probable velocity $u=\sqrt{2k_{\rm B}T/m}$ and given by
\begin{equation}
N(v){\rm d}v=\frac{N_0}{\sqrt{\pi}u} \exp ({-v^2/u^2}){\rm d}v.
\end{equation}
The density matrix elements $\rho_{21}$, $\rho_{31}$ and $\rho_{41}$ represent the coherences between the ground state $\ket{1}$ and excited states $\ket{2}$, $\ket{3}$ and $\ket{4}$ respectively. The ratio $\rho_{31}/\rho_{21}$ between the coherences is 

\begin{equation}
\frac{\rho_{31}}{\rho_{21}}=-\frac{\ii\Omega_{\rm c}}{2} \left[\frac{\Gamma_3}{2}+\ii(k_{\rm p}-k_{\rm c})v+\frac{\Omega_{\rm d}}{2}\frac{\rho_{41}}{\rho_{31}} \right]^{-1},
\end{equation}
where the coherence ratio $\rho_{41}/\rho_{31}$ is 
\begin{equation}
\frac{\rho_{41}}{\rho_{31}}=-\frac{\ii\Omega_{\rm d}}{2}\left[\gamma_{\rm c}-\ii\Delta_{\rm c}+\ii(k_{\rm p}-k_{\rm c}-k_{\rm d})v\right]^{-1},
\end{equation}
$\Delta_{\rm c}$ is the coupling beam detuning from resonance and $\gamma_{\rm c}$ is the dephasing of the Rydberg state due to transit time broadening and laser intensity variations. $\Gamma_2$ and $\Gamma_3$ are the decay rates of states $\ket{2}$ and $\ket{3}$ respectively. The total complex susceptibility $\chi_{\rm tot}$ is given by integrating Eq. (1) over all velocity classes. The total absorption coefficient is given by
\begin{equation}\label{eq:eq2}
\alpha_{\rm tot}=k_{\rm p}{\rm Im}[\chi_{\rm tot}].
\end{equation}

Fig.~3 (a), (b) and (c) show the absorption coefficients per velocity class from the simple four-level theoretical model for $\Omega_{\rm d}$/$\Omega_{\rm c}$ = 0.6, 1.2 and 2.5 respectively as a function of coupling field detuning $\Delta_{\rm c}$ and atomic velocity $v$. We can see that, for different values of $\Omega_{\rm d}/\Omega_{\rm c}$, the gradient of the line defining the frequency of maximum absorption rotates from positive (Fig.~3 (a)) to infinity (Fig.~3 (b)) to negative (Fig.~3 (c)). Significantly, at the optimum ratio of $\Omega_{\rm d}/\Omega_{\rm c}$, the resonance becomes velocity insensitive for $\Delta_{\rm c}=0$~MHz. As a result, more atoms contribute to the signal at the optimum ratio and the absorptive resonance reaches its maximum magnitude, as shown in Fig.~3 (d-f). This occurs because the Doppler shifts of the atoms are compensated by AC Stark shifts \cite{reyn79}. The theoretical fit curves in Fig.~3 (d-f) are calculated by averaging over all magnetic sublevels \cite{tana11}.

As $\Omega_{\rm d}$ increases, the energy separation between the two dressed states $\ket{\rm a^+}$ and $\ket{\rm a^-}$ increases, resulting in the increase in frequency separation between the two EIT resonances as shown by Fig.~4. This splitting scales with the square root of power and has a minimum splitting of approximately 26~MHz which is determined by the wavevectors and Rabi frequencies of the laser fields \cite{tana12}.

To understand the optimum ratio of $\Omega_{\rm d}/\Omega_{\rm c}$, we consider the Hamiltonian of the bare state system ${\cal H}$ given by
\begin{equation}
{\cal H}=\left[ \begin{array}{cccc}
0 & \Omega_{\rm p}/2 & 0 & 0\\
\Omega_{\rm p}/2 & \Delta_{\rm 1ph} & \Omega_{\rm d}/2 & 0\\
0 & \Omega_{\rm d}/2 & \Delta_{\rm 2ph} & \Omega_{\rm c}/2\\
0 & 0& \Omega_{\rm c}/2 & \Delta_{\rm c}+\Delta_{\rm 3ph}
\end{array}\right],
\end{equation}
where $\Delta_{\rm 1ph}=-k_{\rm p}v$, $\Delta_{\rm 2ph}=-(k_{\rm p}-k_{\rm d})v$ and $\Delta_{\rm 3ph}=-(k_{\rm p}-k_{\rm d}-k_{\rm c})v$. To calculate the eigenenergy of the dressed state, we diagonalize the Hamiltonian containing the three upper states and consider the eigenenergy around $v=0$ m/s. We can then expand the characteristic equation in powers of $v$ and neglect higher order terms. The resulting condition for a velocity insensitive bright state is
\begin{equation}\label{eq:eq7}
\frac{\Omega_{\rm d}}{\Omega_{\rm c}}=\sqrt{\frac{k_{\rm p}}{k_{\rm c}+k_{\rm d}-k_{\rm p}}}=1.2.
\end{equation}

\begin{figure}[b]
\centerline{{\includegraphics[width=3.25in]{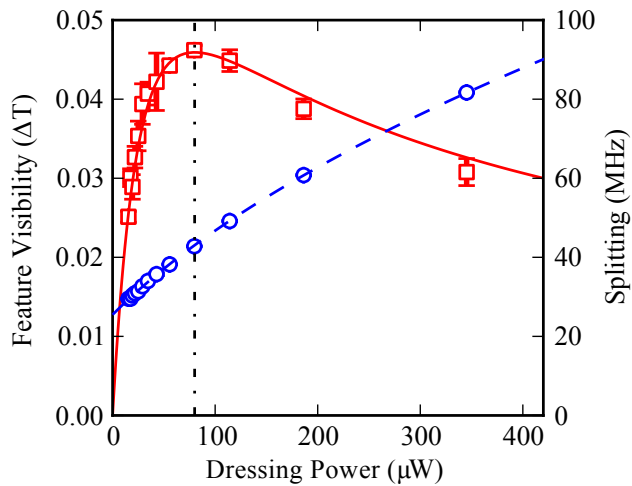}}}
\caption{(color online) Maximum change in transmission (red squares) and transparency splitting (blue circles) as a function of $P_{\rm d}$. The solid (red) line is the theoretical calculation from the full model. The dashed (blue) line is a scaling fit with power. The optimum ratio of $\Omega_{\rm d}/\Omega_{\rm c}$ occurs at P$_{\rm d}$ = 79.7~$\mu$W.}
\end{figure}

Eq. (\ref{eq:eq7}) sets the condition in which the absorption reaches its maximum level as many velocity classes now contribute to the total absorption at $\Delta_{\rm c}$ = 0~MHz. Note that this condition predicts the optimum condition for a single four-level system. However in the real system and our full model, we have many different four-level systems distributed over the magnetic sublevels. The magnitude of the spectral feature as a function of $P_{\rm d}$ for $P_{\rm c}$ = 41.4~mW is shown in Fig~4. The spectral feature has a maximum magnitude at $P_{\rm d}$ = 79.7~$\mu$W. At this point, for the strongest transition, the ratio $\Omega_{\rm d}/\Omega_{\rm c}=1.1$ which is close to the value of $\Omega_{\rm d}/\Omega_{\rm c}=1.2$ predicted by the simple four-level analysis in Eq.~(7). The difference between the actual value and the value predicted by the simple model is due to the inclusion of transitions from all magnetic sublevels.

In summary, we have demonstrated Doppler-compensated Rydberg EIT. We show that there is an optimum ratio of Rabi frequencies for the upper two steps which leads to an enhanced sub-Doppler feature. This work could provide an important step towards single-photon non-linearities in the telecoms wavelength range.


\end{document}